\documentclass[epj]{svjour}
\usepackage{graphics}
\usepackage{epsfig}
\newcommand*{\no}{\noindent}
\newcommand*{\bea}{\begin{eqnarray}}
\newcommand*{\eea}{\end{eqnarray}}
\newcommand*{\be}{\bea}
\newcommand*{\ee}{\eea}
\newcommand*{\pd}{\partial}

\newcommand*{\pdn}{\pd_{\nu}}

\newcommand*{\pref}[1]{(\ref{#1})}

\newcommand*{\mn}{{\mu\nu}}

\newcommand*{\prefr}[2]{(\ref{#1}-\ref{#2})} 
\newcommand*{\nn}{\nonumber}

\begin{document}
\title{On the spectrum of the Faddeev-Popov operator in topological background fields}
\author{Axel Maas}
\institute{Instituto de F\'\i sica de S\~ao Carlos, Universidade de S\~ao Paulo, C.P. 369, 13560-970 S\~ao Carlos, SP, Brazil }
\date{Received: date / Revised version: date}
\abstract{
In the Gribov-Zwanziger scenario the confinement of gluons is attributed to an enhancement of the spectrum of the Faddeev-Popov operator near eigenvalue zero. This has been observed in functional and also in lattice calculations. The linear rise of the quark-anti-quark potential and thus quark confinement on the other hand seems to be connected to topological excitations. To investigate whether a connection exists between both aspects of confinement, the spectrum of the Faddeev-Popov operator in two topological background fields is determined analytically in SU(2) Yang-Mills theory. It is found that a single instanton, which is likely irrelevant to quark confinement, also sustains only few additional zero-modes. A center vortex, which is likely important to quark confinement, is found to contribute much more zero-modes, provided the vortex is of sufficient flux. Furthermore, the corresponding eigenstates in the vortex case satisfy one necessary condition for the confinement of quarks.
 % end of PACS codes
} %end of abstract
\maketitle
\section{Introduction}\label{intro}

The confinement of colored objects in QCD is still a challenging problem, although much progress has been made in its understanding during the last few years \cite{Alkofer:2000wg,Greensite:2003bk}. This progress has been made along two directions, which seem to be quite different at first sight. One is a confinement scenario based on topological defects. The other is based on the properties of field configuration space, in the framework of the Gribov-Zwanziger scenario. It is yet unclear in which way both aspects are connected. The work presented here is aimed at an investigation of a possible link, based on the following set of observations.

The confinement of quarks is usually observed by the existence of an - up to string breaking - linearly rising quark-anti-quark potential. Already more than thirty years ago it was conjectured that topological excitations could be responsible for this behavior. Today, substantial evidence in favor of this scenario exists \cite{Greensite:2003bk}. This would be a very attractive realization, because such excitations can carry topological charge and hence be brought in contact with the spectrum of the Dirac operator via the Atiyah-Singer index theorem \cite{Atiyah:1968mp} and thus to chiral symmetry breaking by the Banks-Casher relation \cite{Banks:1979yr}.

However, the nature of these excitations or corresponding topological defects is currently not finally resolved, nor is its gauge-dependence fully understood. Two of the most important candidates for such configurations are monopoles \cite{Greensite:2003bk,Mandelstam:1974pi} and center vortices \cite{Greensite:2003bk,'tHooft:1977hy,Reinhardt:2001kf}. Vortices seem to have slightly more attractive features in general \cite{Greensite:2003bk}, and hence will be investigated here.

By construction, a non-interacting random ensemble of vortices yields an area law \cite{DelDebbio:1996mh}. Consequently, it has been shown by several groups independently that removal of center vortices removes quark confinement \cite{Greensite:2003bk,Faber:1999sq}. In addition, this also restores chiral symmetry \cite{Gattnar:2004gx}. Thus, vortices will be used here as an example of a quark-confining topological field configuration.

On the other hand, instantons \cite{Belavin:1975fg,Schafer:1996wv}, which play a role in chiral symmetry breaking and hadro-dynamics \cite{Schafer:1996wv,'tHooft:1976fv}, do likely not contribute directly to quark confinement and will be used here as an example of a quark non-confining topological excitation (but see e.\ g.\ \cite{Negele:2004hs} on the topic of instantons and confinement).

The confinement of gluons is to some extent a more subtle issue. As gluons carry adjoint color charges, string breaking is always present. Thus, it is much more complicated to judge whether gluons are confined or not. An empirical criterion for gluon confinement is the Oehme-Zimmermann super-convergence relation \cite{Oehme:bj}. It states that a gluon is confined if its propagator vanishes at zero momentum. Such a behavior has been observed in Landau gauge in calculations based on Dyson-Schwinger equations \cite{Alkofer:2000wg,vonSmekal:1997is,Zwanziger:2001kw} and renormalization group methods \cite{Pawlowski:2003hq}. Corresponding observations have also been made in Coulomb gauge using variational methods \cite{Reinhardt:2004mm}. Lattice calculations have significant problems in reaching sufficiently large volumes (see e.\ g.\ \cite{Bowman:2004jm}), but there are indications on the largest, but highly asymmetric, lattices available \cite{Oliveira:2004gy}, which support a vanishing gluon propagator. In three dimensions, where a similar behavior is found in Dyson-Schwinger calculations \cite{Zwanziger:2001kw,Maas:2004se}, much larger lattices are possible. There, more substantial evidence has been found in favor of such an infrared behavior of the gluon propagator \cite{Cucchieri:2003di}.

This behavior is predicted by the Gribov-Zwanziger confinement scenario \cite{Zwanziger:2001kw,Gribov:1977wm,Zwanziger:2003cf,Zwanziger:2002ia}. The scenario starts from the Gribov problem \cite{Gribov:1977wm,Singer:dk}: Due to the gauge freedom in QCD, not all possible field configurations are independent. It is therefore necessary to restrict the configuration space. Usual local conditions like Landau gauge are only sufficient in perturbation theory. Gribov proposed to restrict further to the first Gribov horizon, a compact region around the origin, hence including perturbation theory. This is the domain in field space where the Faddeev-Popov operator\footnote{As the topological field configurations are defined in Euclidean space, it will be used exclusively here.}
\be
M^{ab}=-\pd_\mu(\pd_\mu\delta^{ab}+gf^{abc}A_\mu^c)\label{fp}
\ee
\no is positive. At the boundary, a zero eigenvalue appears, which becomes negative outside the first Gribov region. Here $g$ is the gauge coupling and $f^{abc}$ are the structure constants of the gauge group. This is not sufficient, as gauge copies exist inside this region \cite{vanBaal:1997gu}. It is thus necessary to restrict even more to the fundamental modular region \cite{Zwanziger:1993dh}, which has in part a common boundary with the first Gribov region. Up to now, no local characterization of the fundamental modular region exists. A non-local possibility \cite{Zwanziger:2003cf} for such a characterization is to choose the configuration on a gauge orbit which absolutely minimizes
\be
\int d^4x (A_\mu^a- A_\mu^{a(n)})(A_\mu^{a}-A_\mu^{a(n)}).\label{minfunc}
\ee
\no Herein $A_\mu^{a(n)}$ is a $n$-instanton configuration to include fields in different instanton sectors.

The Gribov-Zwanziger scenario utilizes the entropy argument that the volume of an infinite dimensional volume is concentrated at its border. Thus, the configurations at the common boundary should dominate the partition sum \cite{Zwanziger:2003cf}. Hence, configurations with a vanishing determinant of the Faddeev-Popov operator \pref{fp} dominate the infrared, and the spectrum of this operator should have an enhancement of zero-modes or near-zero modes compared to the vacuum case.

In Landau gauge, the expectation value of the Faddeev-Popov operator is the inverse ghost propagator. Thus, the scenario predicts an infrared enhanced ghost propagator. It can be shown quite generally that this prediction is correct \cite{Watson:2001yv}; it has also been confirmed in many calculations \cite{vonSmekal:1997is,Zwanziger:2001kw,Pawlowski:2003hq,Zwanziger:2003cf}, including lattice ones \cite{Langfeld:2002dd,Sternbeck:2005tk}. The connection between such an enhancement of the spectrum and an infrared divergent ghost propagator has also been shown explicitly in lattice calculations \cite{Sternbeck:2005vs}. Furthermore, the scenario predicts that the infrared behavior of Yang-Mills theory is dominated by the gauge-fixing part of the action \cite{Zwanziger:2003cf}, which has been confirmed in Dyson-Schwinger calculations \cite{vonSmekal:1997is}. In addition, the properties of the ghost-gluon vertex have been found to agree very well with this scenario \cite{Zwanziger:2003cf,Schleifenbaum:2004id}. Similar relations hold in Coulomb gauge, which have also been confirmed in functional and lattice calculations \cite{Reinhardt:2004mm,Greensite:2004ur}.

Corresponding predictions are also given by a scenario proposed by Kugo and Ojima \cite{Kugo:gm}, which is based on the BRST symmetry. Especially, at least in Landau gauge, an enhancement of the ghost propagator is predicted \cite{Kugo:1995km}, and therefore an enhancement of additional zero-modes in the spectrum of the Faddeev-Popov operator. It is, however, not yet clear, if a connection between both scenarios exist.

These observations let it appear very promising to investigate analytically which type of connection between quark-confining topological field configurations and an enhancement of zero or near-zero modes of the Faddeev-Popov operator exists. This is also motivated by independent results on this topic in lattice gauge theory \cite{Greensite:2004ur}. As a cross-check, it is interesting to study a configuration like an instanton, which is probably not directly involved in quark confinement.

This will be the content of this work: In section \ref{vac}, the spectrum of the Faddeev-Popov operator in the vacuum will be discussed, in order to fix notations, for comparison, and for the sake of completeness. The spectrum in a one-instanton background will be determined in section \ref{inst} and for in a one-center-vortex background in section \ref{vor}. In section \ref{ffunc}, a necessary criterion for confinement will be tested for the solution in the vortex case. The results and some possible implications will be discussed in section \ref{dc}.

\section{Vacuum}\label{vac}

In the vacuum the Faddeev-Popov operator \pref{fp} is just the negative of the 4-dimensional Laplacian times a unit matrix in color space,
\bea
M^{ab}&=&-\delta^{ab}\pd^2=-\delta^{ab}\left(\frac{\pd^2}{\pd x^2}+\frac{\pd^2}{\pd y^2}+\frac{\pd^2}{\pd z^2}+\frac{\pd^2}{\pd t^2}\right)\nn\\
&=&-\delta^{ab}\Big(\frac{1}{r^3}\pd_r r^3\pd_r+\frac{1}{r^2\sin^2\eta}\pd_\eta\sin^2\eta\pd_\eta\nn\\
&&+\frac{1}{r^2\sin^2\eta}\Big(\frac{1}{\sin^2\theta}\pd^2_\phi+\frac{1}{\sin\theta}\pd_\theta\sin\theta\pd_\theta\Big)\Big)\label{lphc}\\
&=&-\delta^{ab}\left(\frac{1}{r}\pd_r r\pd_r+\frac{1}{r^2}\pd_\theta^2+\frac{1}{\rho}\pd_\rho\rho\pd_\rho+\frac{1}{\rho^2}\pd_\eta^2\right).\label{bipol}
\eea
\no The second expression is given in hyper-spherical coordinates
\bea
r_\mu&=&(x,y,z,t)^T\\
&=&(r\cos\phi\sin\theta\sin\eta,r\sin\phi\sin\theta\sin\eta,\nn\\
&&r\cos\theta\sin\eta,r\cos\eta)^T\nn,
\eea
\no where $\phi$ ranges in $[0,2\pi)$ and $\theta$ and $\eta$ range in $[0,\pi)$. The third expression is in bi-polar coordinates, essentially a set of two polar coordinate systems,
\be
r_\mu=(r\cos\theta,r\sin\theta,\rho\cos\eta,\rho\sin\eta)^T.\label{bipolar}
\ee
\no In the vacuum the eigenmode equation is separable and thus directly solvable.

In Cartesian coordinates, the eigenmodes, solutions of the eigenequation
\be
M^{ab}\phi^b=\omega^2\phi^a\nn
\ee
\no of the vacuum Faddeev-Popov operator are plane waves $C^a\exp(ik_\mu r_\mu)$, and the spectrum is the continuum of positive real numbers $\omega^2$ with $\omega^2=k_\mu k_\mu$. Note that in the vacuum the operator is positive definite. As these are free states, they are non-normalizable and correspond to scattering states in quantum mechanics. In the vacuum the colors always decouple, and there is always a set of $N_c^2-1$ independent solutions. The only zero-modes are trivial, constant functions, of which also $N_c^2-1$ linearly independent ones exist.

In hyper-spherical coordinates, the eigenmodes can be decomposed into partial waves with three independent integer angular quantum numbers $n$, $l$, and $m$ \cite{Cucchieri:1995yd}. The eigenmodes are then given by
\be
\phi^a(r_\mu)=C^a\frac{J_n(\omega r)}{\omega r}\frac{P^{l+1/2}_{n-1/2}(\cos\eta)}{\left(\cos^2\eta-1\right)^{1/4}}Y^l_m(\theta,\phi),\nn
\ee
\no where $J$ are the Bessel functions and $P$ are the associated Legendre functions of the first kind. $Y^l_m$ are the ordinary spherical harmonics. A non-vanishing zero-mode is only possible for $n=l=m=0$. These are again constant solutions.

However, further zero-modes, e.\ g.\ of the form $C^a/r^2$, also exist. These will not be admitted as they are in general more singular than the gauge field configuration. This will also be required when treating non-zero gauge field configurations below. This will imply that in the cases treated here any admissible solution may not diverge at spatial infinity.

In bi-polar coordinates, the problem separates into two two-dimensional sets, one for each polar coordinate system. As both angular coordinates are $2\pi$-periodic, the solution is
\be
\phi^a=C^aJ_{|n|}\left(\omega\sqrt{1-s^2}\rho\right)J_{|m|}(\omega s r)\exp(i(m\theta+n\eta)).\label{fsol}
\ee
\no $J_n$ are again the Bessel functions. The angular quantum numbers $m$ and $n$ are positive and negative integers including zero and can be chosen independently for each color. The continuous variable $s$, which ranges in $[0,1]$, ``shifts the eigenvalue'' between both coordinate sets. Again there are only trivial zero-modes, all constants: For $\omega=0$, only $n=m=0$ leads to non-vanishing values of the Bessel functions.

\section{Instanton}\label{inst}

\subsection{Field configuration}\label{finst}

The simplest case of a topological field configuration in Yang-Mills theory is an instanton\footnote{The remaining part of the article is restricted to SU(2) Yang-Mills theory.}. These are given as algebra elements by \cite{Bohm:2001yx}
\bea
A_\mu&=&\frac{2}{r^2+\lambda^2}\tau_\mn r_\nu\nn\\
\tau_\mn&=&\frac{1}{4i}(\tau_\mu\bar\tau_\nu-\tau_\nu\bar\tau_\mu)\nn\\
\tau_\mu&=&(i\vec\tau,1)\nn\\
\bar\tau_\mu&=&(-i\vec\tau,1),\nn
\eea
\no where $\lambda$ characterizes the size of the instanton and $\tau^i$ are the Pauli matrices. The corresponding gauge fields are given by the regular functions
\be
A_\mu^a=\frac{1}{g}\frac{2}{r^2+\lambda^2}r_\nu \zeta^a_{\nu\mu}.\label{ifield}
\ee
\no The constant real matrices $\zeta^a$ (the 't Hooft tensors) form the algebra
\bea
[\zeta^a,\zeta^b]&=&2f^{abc}\zeta^c\label{ga1}\\
\{\zeta^a,\zeta^b\}&=&-\delta^{ab}.\label{ga2}
\eea

The instanton fields are transverse, $\pd_\mu A^a_\mu=0$, and thus admissible in Landau gauge. Furthermore, they are part of the fundamental modular region as they trivially minimize \pref{minfunc}. Hence, the Faddeev-Popov operator should have only positive or zero eigenvalues, which will be confirmed below.

\subsection{Analytical treatment}\label{ainst}

Due to the structure of \pref{ifield} and the transversality of the field, the eigenvalue equation of the Faddeev-Popov operator in an instanton field can be written as
\be
\pd^2\phi^a+f^{abc}\frac{2}{r^2+\lambda^2}r_\mu \zeta^b_\mn \pdn\phi^c=-\omega^2\phi^a.\label{icomp}
\ee
\no Neither do the different colors decouple nor is the problem fully separable anymore. As it is possible to express distances $r$ in dimensionless variables $r/\lambda$, only the dimensionless quantity $\omega\lambda$ is left as an independent parameter.

First of all, due to the transversality of the instanton field, the three trivial constant zero-modes still exist.

To solve the set of equations \pref{icomp}, the first step is to note that the expressions $r_\mu \zeta^a_\mn\pd_\nu$ are angular momentum operators. The eigenvalue equation can then be rewritten as
\bea
0&=&(\pd^2+\omega^2)\phi^1\label{iq1}\\
&&+\frac{2}{i(r^2+\lambda^2)}((L_{34}+L_{12})\phi^2+(-L_{24}+L_{13})\phi^3)\nn\\
0&=&(\pd^2+\omega^2)\phi^2\label{iq2}\\
&&+\frac{2}{i(r^2+\lambda^2)}(-(L_{34}+L_{12})\phi^1+(L_{14}+L_{23})\phi^3)\nn\\
0&=&(\pd^2+\omega^2)\phi^3\label{iq3}\\
&&+\frac{2}{i(r^2+\lambda^2)}((L_{24}-L_{13})\phi^1-(L_{14}+L_{23})\phi^2)\nn\\
L_{\alpha\beta}&=&i(x_\alpha\pd_\beta-x_\beta\pd_\alpha)\nn.
\eea
\no The angular momentum operators satisfy the commutation relations
\be
[L_{\alpha\beta},L_{\gamma\delta}]=i(\delta_{\beta\gamma}L_{\alpha\delta}+\delta_{\alpha\delta}L_{\beta\gamma}+\delta_{\beta\delta}L_{\gamma\alpha}+\delta_{\alpha\gamma}L_{\delta\beta}).\nn
\ee
\no This implies that any solution which only depends on $r$ does not feel the presence of the instanton, and is the same solution as in the vacuum. Thus, there exists a continuous positive spectrum, as expected. Furthermore, the Laplacian can be written as \cite{Cucchieri:1995yd}
\be
\pd^2=\frac{1}{r^3}\pd_r r^3\pd_r-\frac{1}{r^2}\sum_{1\le\alpha<\beta\le 4} L_{\alpha\beta}L_{\alpha\beta}\label{laplace}.
\ee
\no As only three particular linear combinations of the six angular momentum operators appear, it is possible to define the set of operators
\bea
L^1&=&\frac{1}{2}(L_{14}+L_{23})\nn\\
L^2&=&\frac{1}{2}(L_{13}-L_{24})\nn\\
L^3&=&\frac{1}{2}(L_{34}+L_{12})\nn,
\eea
\no which form a three-dimensional spin-algebra
\be
[L^a,L^b]=if^{abc}L^c.\nn
\ee
\no In addition, explicit calculation shows that
\bea
\vec L^2&=&(L^1)^2+(L^2)^2+(L^3)^2\nn\\
&=&\frac{1}{4}\sum_{1\le\alpha<\beta\le 4} L_{\alpha\beta}L_{\alpha\beta}\nn\\
&&+\frac{1}{2}(L_{14}L_{23}-L_{13}L_{24}+L_{12}L_{34}).\nn
\eea
\no The second term vanishes. Thus, the angular part of the Laplacian operator can be replaced by $4\vec L^2$.
\be
\pd^2=\frac{1}{r^3}\pd_r r^3\pd_r-\frac{4\vec L^2}{r^2}.\nn
\ee
\no Hence, the problem can be separated in a radial part and an angular part. The latter can be characterized by two independent quantum numbers $l$ and $m$. Selecting $L^3$ as ``quantization'' axis\footnote{This is the axis characterizing rotations in the 1-2 and 3-4 planes around the same angle. In fact bi-polar coordinates would be the natural way to solve this problem explicitly in terms of $r^2+\rho^2$, $r/\rho$ and $\theta+\eta$. However, this will be not necessary for the purpose at hand.}, the eigenfunctions depend on $l$, denoting an ``orbital angular momentum'', which can be integer or half-integer. The ``magnetic'' eigenstates are then counted by $m$. In principle, each color could have its own angular momentum $l$, but in the equations \prefr{iq1}{iq3} no operator appears which can change $l$. So to compensate the angular dependencies, each color has to have the same $l$-value for a solution, and $\vec L^2$ can be replaced by $l(l+1)$ as its eigenvalue.

Thereby each of the equations splits in two parts,
\bea
0&=&\left(\frac{1}{r^3}\pd_r r^3\pd_r-\frac{4l(l+1)}{r^2}+\omega^2\right)\phi^1\nn\\
&&+\frac{4}{i(r^2+\lambda^2)}\left(L^3\phi^2+L^2\phi^3\right)\nn\\
0&=&\left(\frac{1}{r^3}\pd_r r^3\pd_r-\frac{4l(l+1)}{r^2}+\omega^2\right)\phi^2\nn\\
&&+\frac{4}{i(r^2+\lambda^2)}\left(-L^3\phi^1+L^1\phi^3\right)\nn\\
0&=&\left(\frac{1}{r^3}\pd_r r^3\pd_r-\frac{4l(l+1)}{r^2}+\omega^2\right)\phi^3\nn\\
&&+\frac{4}{i(r^2+\lambda^2)}\left(-L^2\phi^1-L^1\phi^2\right)\nn.
\eea
\no Using a matrix representation for the angular momentum operators, this system can be written as a matrix equation
\be
(\underline 1 D_r+\underline L_I)\vec\phi=0.\label{eveq}
\ee
\no The dimensionality of the vector $\vec \phi$ is $3(2l+1)$, representing the three colors and the $(2l+1)$ independent $m$-quantum numbers. $D_r$ collects the radial derivatives and terms after multiplication with $r^2+\lambda^2$. $\underline 1$ denotes the unit matrix in this space and $\underline L_I$ is the hermitian matrix
\be
\underline L_I=\frac{4}{i}\pmatrix{0 & L^3 & L^2 \cr -L^3 & 0 & L^1 \cr -L^2 & -L^1 & 0 \cr}.\label{matrix}
\ee
\no Expanding the solution $\vec\phi$ in the eigenbasis $\{\vec l_i\}$ of $\underline L_I$ as $\sum \phi_{li}(r)\vec l_i$ completely decouples the angular and radial parts, leading to the radial equation
\be
(D_r+c_i)\phi_{li}(r)=0.\label{radialeq}
\ee
\no Here, $c_i$ are the eigenvalue of the corresponding eigenvector. The eigenvalue problem is a typical spin problem. In general, there are $3(2l+1)$ eigenvalues, which can be either positive or negative or eventually zero. Note that in the case $l=0$ the solution are functions dependent only on $r$, as already discussed above. Therefore, only $l\ge 1/2$ will be regarded here. The eigenvectors will not be needed to determine the spectrum, but could be obtained in a direct way. However, to solve the radial equation \pref{radialeq}, the eigenvalues have to be determined. A close inspection of the eigenvalue problem yields the characteristic polynomial of \pref{matrix} as
\be
(c-4)^{2l+1}(c+4l)^{2l+3}(c-4(l+1))^{2l-1}=0.\nn
\ee
\no Hence there are only three different eigenvalues $c_i$: $4$, $-4l$, and $4(l+1)$ with multiplicities $2l+1$, $2l+3$, and $2l-1$, respectively. The radial solution does not depend on the multiplicity index and $\phi_{li}$ can be replaced by $\phi_{lc}$. With this, the radial equation can finally be written down as
\be
0=\frac{1}{r^3}\pd_r r^3\pd_r\phi_{lc}+\left(\omega^2-\frac{4l(l+1)}{r^2}+\frac{c}{r^2+\lambda^2}\right)\phi_{lc}.\label{req}
\ee
\no The only difference between the different colors are then their components along the quantization axis and thus their composition in terms of the different magnetic eigenstates: The different colors are rotated with respect to each other.

A first step to solve the ordinary differential equation for the radial part \pref{req} is to investigate the asymptotic properties in order to study the existence of admissible solutions. At small $r$, the only relevant part is the angular part, leading to the simpler equation
\be
0=\pd_r^2\phi_{lc}+\frac{3}{r}\pd_r\phi_{lc}-\frac{4l(l+1)}{r^2}\phi_{lc}.\nn
\ee
\no This equation is independent of $c$ and $\omega^2$. It is solved by
\be
\phi(r)=_{r\to0}C_1 r^{2l}+C_2 r^{-2(l+1)},\nn
\ee
\no with $C_1$ and $C_2$ integration constants. Therefore, there exists only one solution which is regular at the origin, and there is no more than one non-singular solution for the differential equation \pref{req}. The homogeneity of the differential equation then implies that the only remaining integration constant is an over-all normalization factor.

At large distances only the part containing the eigenvalue in \pref{req} is relevant as long as $\omega^2$ is non-vanishing. Thus, the more simple differential equation
\be
\pd_r^2\phi_{lc}+\frac{3}{r}\pd_r\phi_{lc}+\omega^2\phi_{lc}=0\nn
\ee
\no has to be treated. This is just the equation for $l=0$, which is solved by Bessel functions of the first and the second kind. In this case the second solution is also admissible, as the behavior at the origin is not relevant, yielding
\be
\phi^a_{cl}\to D_1\frac{J_n(\omega r)}{r}+D_2\frac{Y_n(\omega r)}{r}.\label{infsol}
\ee
\no $D_1$ and $D_2$ are new integration constants. In principle, \pref{infsol} allows to construct finite solutions for any positive or negative $\omega^2$ by an appropriate choice of $D_1$ and $D_2$. Thus, besides showing that the long-range regular solutions will be of type $\cos(r/\lambda+\delta)/r^{3/2}$, \pref{infsol} does not provide further information. It is therefore necessary to obtain an explicit solution for the differential equation \pref{req}. Making the ansatz $\phi_{lc}=r^{2l}w_{lc}(r)$, and expanding then $w_{lc}(r)$ in powers of $r/\lambda$, this is straightforward. The general solution for any $\omega^2$, $l$, and $c$ is given by
\bea
\phi_{lc}&=&r^{2l}\sum_{n=0}^{\infty}a_n\left(-\frac{r^2}{\lambda^2}\right)^n\label{srep}\\
a_{-1}&=&0\nn\\
a_{0}&=&D\nn\\
a_{n}&=&\frac{\omega^2\lambda^2(a_{n-1}-a_{n-2})}{4n(n+1)+8ln}\nn\\
&&+\frac{((4n+8l)(n-1)+c)a_{n-1}}{4n(n+1)+8ln}\nn.
\eea
\no Here $D$ is the free overall normalization factor. Unfortunately, this series is similar to the hyper-geometric series, and its convergence\footnote{To be precise: As this is a solution to an initial value problem for a smooth differential equation, the series is guaranteed to exist and converge for any finite $r$. This, however, does not prevent a divergence for $r\to\infty$.} cannot be checked easily for $r>\lambda$. It is therefore necessary to perform a numerical analysis, which will be done in section \ref{rinst}.

\subsection{Results}\label{rinst}

Beforehand, it is worthwhile to treat the case $\omega^2=0$ explicitly. The equation
\be
0=\pd_r^2\phi^a_{lc}+\frac{3}{r}\pd_r\phi^a_{lc}+\left(-\frac{4l(l+1)}{r^2}+\frac{c}{r^2+\lambda^2}\right)\phi^a_{lc}=0\nn
\ee
\no is directly solvable, and yields hyper-geometric functions
\bea
\phi_{lc}(r)&=&C_1\left(\frac{\lambda^2}{r^2}\right)^{1+l}\; _2 F_1\Big(-\frac{1}{2}-l-\frac{1}{2}\sqrt{1-c+4l+4l^2},\nn\\
&&-\frac{1}{2}-l+\frac{1}{2}\sqrt{1-c+4l+4l^2},-2l,-\frac{r^2}{\lambda^2}\Big)\nn\\
&&+C_2\left(\frac{r^2}{\lambda^2}\right)^{l}\; _2 F_1\Big(\frac{1}{2}+l-\frac{1}{2}\sqrt{1-c+4l+4l^2},\nn\\
&&\frac{1}{2}+l+\frac{1}{2}\sqrt{1-c+4l+4l^2},2+2l,-\frac{r^2}{\lambda^2}\Big)\nn.
\eea
\no As $_2F_1(\alpha,\beta,\gamma,0)=1$, the first of these solutions is singular at the origin and can be dismissed. The second solution is regular. Thus, it remains to investigate its long distance behavior. Using an identity for the hyper-geometric functions \cite{Gradstein:1981}, the expression can be rewritten as
\bea
&&\left(\frac{r^2}{\lambda^2}\right)^{l}\left(1+\frac{r^2}{\lambda^2}\right)^{-(\frac{1}{2}+l-\frac{1}{2}\sqrt{1-c+4l+4l^2})}\nn\\
&&\times_2 F_1\Big(\frac{1}{2}+l-\frac{1}{2}\sqrt{1-c+4l+4l^2},\nn\\
&&\frac{3}{2}+l-\frac{1}{2}\sqrt{1-c+4l+4l^2},2+2l,\frac{\frac{r^2}{\lambda^2}}{\frac{r^2}{\lambda^2}+1}\Big).\nn
\eea
\no For $r\to\infty$, the hyper-geometric function is then a constant for all eigenvalues. The pre-factor is finite at spatial infinity only for $l=1$ and $c=8$ with multiplicity one and $l=1/2$ and $c=4$ with multiplicity two. In these cases the zero-modes can be given in closed form and read for $l=1/2$
\be
\phi_{\frac{1}{2}\;4}(r)=2D\frac{-\frac{r^2}{\lambda^2}+\left(1+\frac{r^2}{\lambda^2}\right)\ln\left(1+\frac{r^2}{\lambda^2}\right)}{\frac{r^3}{\lambda^3}}\nn
\ee
\no and for $l=1$
\be
\phi_{1\;8}(r)=3D\frac{\frac{r^4}{\lambda^4}+2\frac{r^2}{\lambda^2}-2(1+\frac{r^2}{\lambda^2})\ln\left(1+\frac{r^2}{\lambda^2}\right)}{\frac{r^4}{\lambda^4}}.\nn
\ee
\no No further zero-modes exist for larger $l$ or for any other eigenvalue $c$. Hence, the instanton sustains six zero-modes in total, the three trivial ones at $l=0$, two for $l=1/2$ and one for $l=1$.

\begin{figure}
\epsfig{file=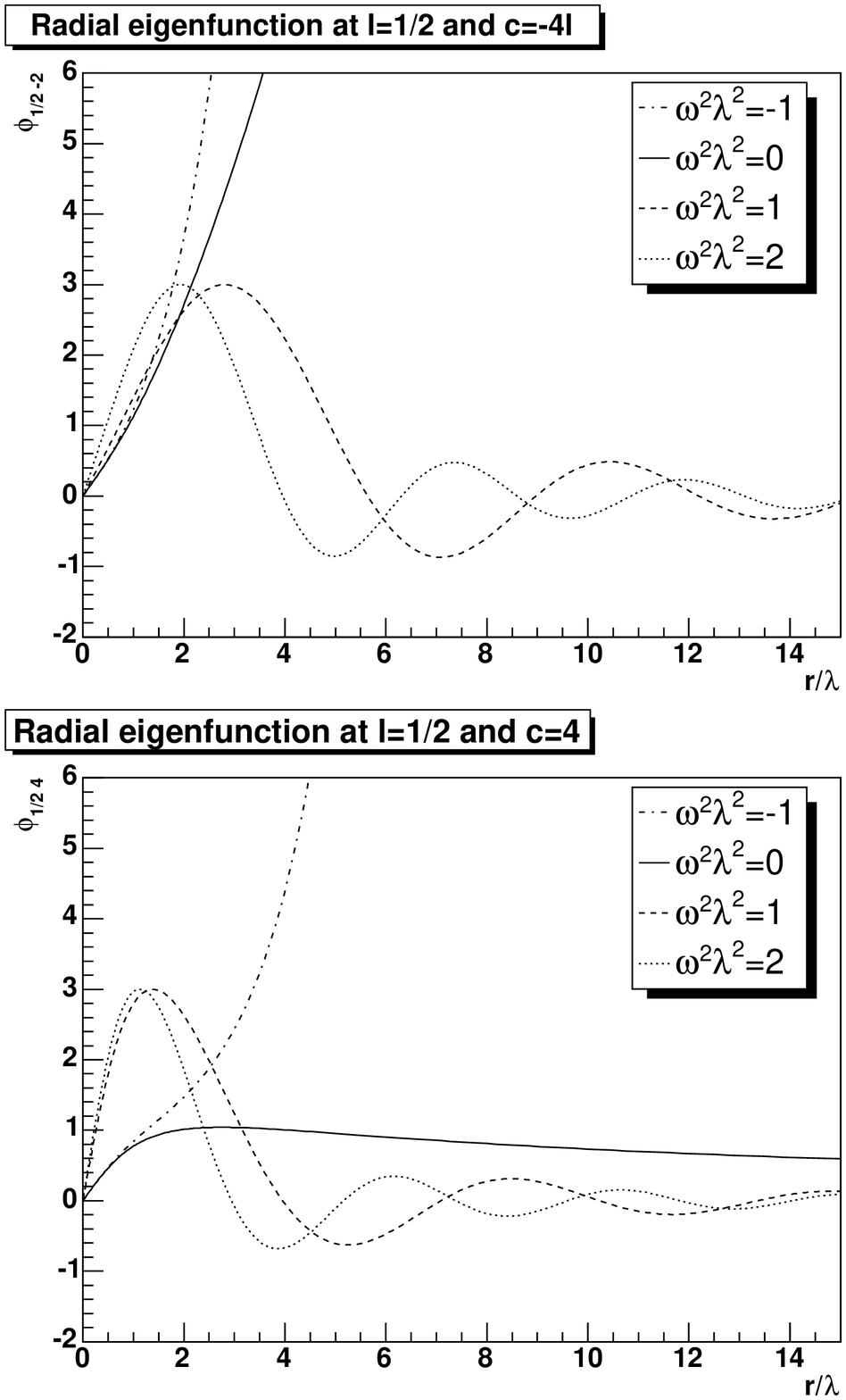,width=\linewidth}
\caption{The radial eigenfunctions $\phi_{lc}$ for $l=1/2$ and the two different $c$ values. For better visualization, positive $\omega^2$-solutions have been normalized so that their maximum is $3$, while modes with $\omega^2\lambda^2\le 0$ have been normalized so that $\phi_{lc}/r^{2l}|_{r=0}=1$.}\label{figl0.5}
\end{figure}

\begin{figure}
\epsfig{file=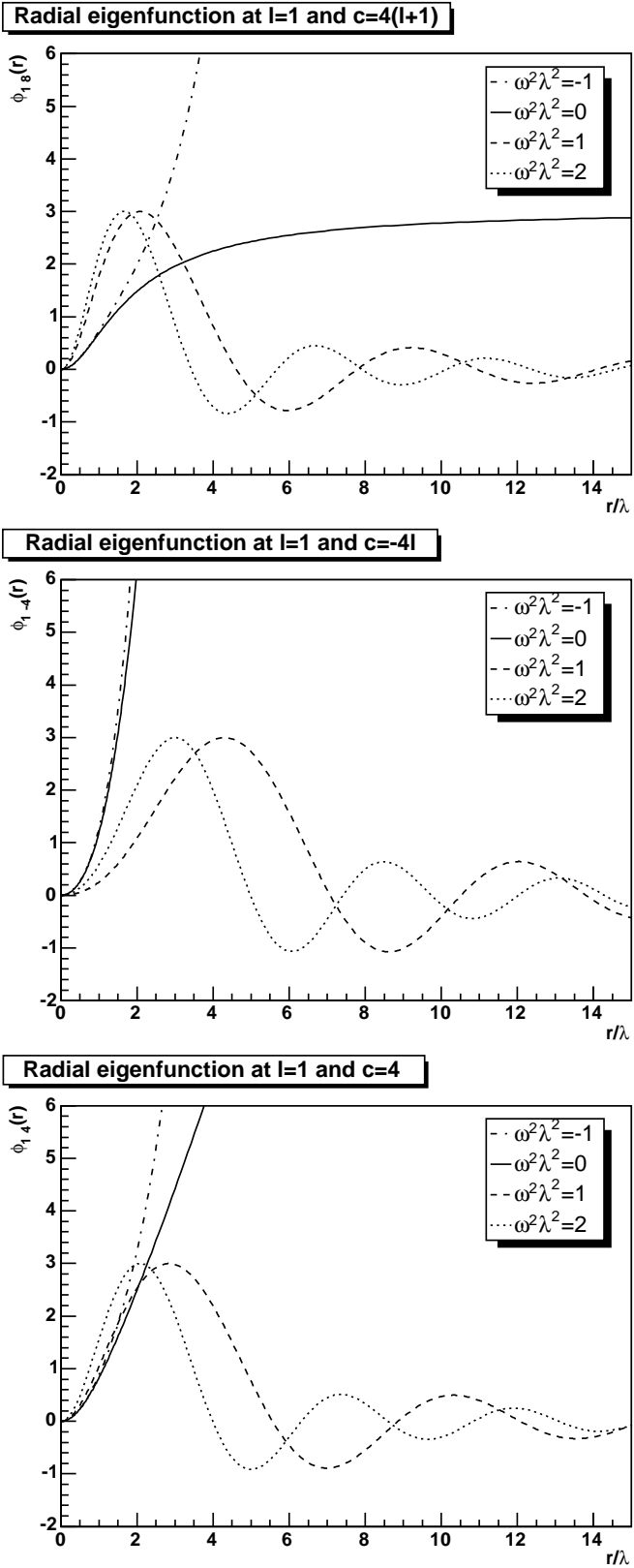,width=\linewidth}
\caption{The radial eigenfunctions $\phi_{lc}$ for $l=1$ and the three different $c$ values. For better visualization, positive $\omega^2$-solutions have been normalized so that their maximum is $3$, while modes with $\omega^2\lambda^2\le 0$ have been normalized so that $\phi_{lc}/r^{2l}|_{r=0}=1$.}\label{figl1}
\end{figure}

These results can be supported by numerical calculations. A selection of numerical results\footnote{These were obtained by summing up the series explicitly for $r/\lambda< 1$. At some point $r_s$, the series was used to find initial values for the differential equation, which was then solved by a Cash-Karp Runge-Kutta algorithm \cite{Press:1997}. This was necessary due to explicit divergencies of the angular term in the differential equation at 0. Otherwise the numerical solution could have been obtained by the integration method alone.} is shown in figure \ref{figl0.5} for $l=1/2$ and in figure \ref{figl1} for $l=1$ for the possible values of $c$ and various values of $\omega^2\lambda^2$. Calculations at larger $l$ or other values of $\omega^2\lambda^2$ confirm these findings. No further zero-modes exist.

\subsection{Singular gauge}

It is an interesting question, to which extent these findings are gauge dependent. In general, different gauges lead to much more complicated differential equations, especially if the field configuration is no longer transverse afterwards. In the case of the instanton, it is possible to check at least one other gauge. Performing the gauge transformation \cite{Bohm:2001yx}
\be
G(x)=\frac{\tau_\mu r_\mu}{r},\nn
\ee
\no the instanton field configuration is given by \cite{Bohm:2001yx}
\bea
A_\mu&=&\frac{2\lambda^2}{r^2(r^2+\lambda^2)}\bar\tau_\mn r_\nu\nn\\
\bar\tau_\mn&=&\frac{1}{4i}(\bar\tau_\mu\tau_\nu-\bar\tau_\nu\tau_\mu)\nn.
\eea
\no This gauge transformation has the property to move the non-trivial behavior from infinity to a finite region, at the cost of a small-distance divergence. In a sense, this gauge transformation is one of the most severe changes which can be done to the global, and thus long-range and confinement-relevant, properties of a gauge field configuration. Therefore, it is an important test of the results found above.

The problem can be treated along the same lines as before, but as the field configuration now diverges as $1/r^2$ at the origin, corresponding eigenfunctions have to be admitted. It is then still possible to separate the angular structure, and the equations turn out to be the same as \pref{eveq}, except for a different quantization axis (which nonetheless can be called $L^3$). In addition, an additional factor of $r^2/\lambda^2$ appears in the radial part. Thus, the same angular solutions are found, and the radial equation takes the form
\be
0=\frac{1}{r^3}\pd_r r^3\pd_r\phi_{lc}+\left(\omega^2-\frac{4l(l+1)}{r^2}+\frac{c\lambda^2}{r^2(r^2+\lambda^2)}\right)\phi_{lc}\nn
\ee
\no with the same values of $c$ as before. The solutions of this equation differ qualitatively from their previous counterparts only by an admitted divergence at zero, and are given in explicit form for $\omega^2\neq 0$ by
\bea
\phi_{lc}&=&r^{-1+\sqrt{(2l+1)^2-c}}\sum_{n=0}^{\infty}a_n\left(\frac{r^2}{\lambda^2}\right)^n\nn\\
a_{-1}&=&0\nn\\
a_{0}&=&D\nn\\
a_{n}&=&\Big(\Big(\Big(c-\lambda^2\omega^2\Big)-4(n-1)^2\nn\\
&&-2(n-1)\sqrt{(2l+1)^2-c}\Big)a_{n-1}-\lambda^2\omega^2a_{n-2}\Big)\nn\\
&&\times\frac{1}{2n\left(2n+\sqrt{(2l+1)^2-c}\right)}\nn.
\eea
\no There are three non-trivial zero-modes, at exactly the same values of $l$ and $c$, reading
\bea
\phi_{\frac{1}{2}\;4}&=&\frac{D}{2}\frac{1+\left(1+\frac{r^2}{\lambda^2}\right)\ln\left(\frac{r^2}{r^2+\lambda^2}\right)}{\frac{r}{\lambda}}\nn\\
\phi_{1\; 8}&=&D\frac{1+2\frac{r^2}{\lambda^2}+2\left(\frac{r^2}{\lambda^2}+\frac{r^4}{\lambda^4}\right)\ln\left(\frac{r^2}{r^2+\lambda^2}\right)}{\frac{r^2}{\lambda^2}}.\nn
\eea
\no These solutions decay faster than in the regular case, and are even square-integrable and therefore normalizable and localized.

Note that there are now three additional zero-modes at $l=0$, $\phi(r)_{0\; 0}=C^a/r^2$, which previously have been rejected in a non-singular field-configuration. These are also trivial zero-modes, in the sense that they already exist in the vacuum, and the number of non-trivial zero-modes is the same.

Hence, the non-trivial properties of the spectrum is not affected by this gauge transformation. This is not entirely expected, as the Faddeev-Popov operator is gauge-dependent, and the chosen gauge transformation severely affects the long distance properties of the field configuration.

\section{Center Vortex}\label{vor}

\subsection{Field configuration}\label{fvor}

In this section the spectrum of the Faddeev-Popov operator will be studied in the background of a thick, oriented center vortex. The vortex field used here provides a non-trivial Wilson loop in the fundamental representation of the gauge group. The field strength of such a vortex in SU(2) is given by \cite{Diakonov:1999gg} 
\be
A_\eta^a=\delta^{3a}\frac{1}{g}\frac{\mu(\rho)}{\rho}.\nn
\ee
\no All other components vanish\footnote{In principle, it would be possible that there is a non-vanishing component $A_\rho$. Such a component can be removed by choosing a suitable gauge \cite{Diakonov:1999gg}.}. This field is transverse. The function $\mu(\rho)$ varies from zero at $\rho=0$ to zero at $\rho=\infty$ for a flux zero vortex. For higher fluxes, it varies from zero at $\rho=0$ to the flux $2n+1$ at $\rho=\infty$. Herein, $n$ is a positive integer or zero. There is no further specification of this ``profile'' of the vortex. For the results, several different ``profiles'' will be compared. The only assumption made will be that they are smooth.

This type of center vortices are energy minimizing configurations after including first order quantum effects \cite{Diakonov:1999gg}. At the classical level, only instantons are such solutions.

\subsection{Analytical treatment}\label{avor}

The natural coordinate system for this problem are bi-polar coordinates \pref{bipolar}. The three eigenvalue equations for the three colors are then given by
\bea
-\pd^2\phi^1-\frac{\mu}{\rho^2}\pd_\eta\phi^2&=&\omega^2\phi^1\nn\\
-\pd^2\phi^2+\frac{\mu}{\rho^2}\pd_\eta\phi^1&=&\omega^2\phi^2\nn\\
-\pd^2\phi^3&=&\omega^2\phi^3\nn.
\eea
\no The equation for the 3-component decouples and thus is solved by the free solution \pref{fsol}. This implies that $\phi^3$ becomes constant for any zero-mode. It thus remains to solve the coupled equations for the eigenfunctions of colors one and two. For these the behavior in $r-\theta$, in which they are trivial, and $\rho-\eta$ decouple. The equations are then reduced to a two-dimensional problem
\bea
\left(\frac{1}{\rho}\pd_\rho\rho\pd_\rho+\frac{1}{\rho^2}\pd_\eta^2+\omega^2(1-s^2)\right)\phi^1+\frac{\mu}{\rho^2}\pd_\eta\phi^2&=&0\nn\\
\left(\frac{1}{\rho}\pd_\rho\rho\pd_\rho+\frac{1}{\rho^2}\pd_\eta^2+\omega^2(1-s^2)\right)\phi^2-\frac{\mu}{\rho^2}\pd_\eta\phi^1&=&0\nn.
\eea
\no As the angle $\eta$ is $2\pi$-periodic, it is possible to expand the eigenfunctions in a Fourier-series as $\phi^a=\sum_m c_m^a\exp(im\eta)$. This yields
\bea
0&=&\sum_m\Big(\Big(\frac{1}{\rho}\pd_\rho\rho\pd_\rho-\frac{m^2}{\rho^2}+\omega^2(1-s^2)\Big)c_m^1\nn\\
&&+\frac{im\mu}{\rho^2}c_m^2\Big)\exp(im\eta)=0\label{feq1}\\
0&=&\sum_m\Big(\Big(\frac{1}{\rho}\pd_\rho\rho\pd_\rho-\frac{m^2}{\rho^2}+\omega^2(1-s^2)\Big)c_m^2\nn\\
&&-\frac{im\mu}{\rho^2}c_m^1\Big)\exp(im\eta)\label{feq2}.
\eea
\no Only the equations for the same $m$ are coupled, as $\eta$ is arbitrary, and it suffices to solve the equations for each $m$ separately.

At $m=0$ the system decouples, as it is reduced to the free case. Consequently, this yields $c_0^a=C^aJ_0\left(\omega\sqrt{1-s^2}\rho\right)$ as a solution. This implies that $s$ is again constrained to $0\le s\le 1$, which is confirmed by the large $\rho$ solution. Thus, $m$ can be taken non-zero in the following.

Since $\mu$ goes to zero at the origin, the angular term $m^2/\rho^2$ dominates, just as in the case of the instanton. The solution is then similarly given for small\footnote{Small is here determined by the typical scale of the profile function $\mu$. Whenever there is a notion of large and small, this will be with respect to this intrinsic scale, which is only determined if the function $\mu$ is given explicitly.} $\rho$ by
\be
c_m^a\sim A^a\rho^{|m|}+B^a\rho^{-|m|},\label{irsol}
\ee
\no with integration constants $A^a$ and $B^a$. As for all profiles the field configurations vanish at the origin faster than $1/\rho$, the second solution is dismissed. This requires two of the four free (complex) integration constants. As the differential equations are homogenous in the functions, the remaining integration constants are again scale factors.

At very large distances for non-zero $\omega$, the eigenvalue term dominates. The equations decouple and in this case the eigenfunctions will behave, as in the free case, like Bessel-functions. Thus, for finite $\omega$ the influence of the vortex is only relevant at intermediate distances. The situation is different in the case of a zero-mode, and will be treated below.

To make progress with the solution, it is useful to rewrite equations \prefr{feq1}{feq2} with explicit real and imaginary parts. Then the equations for the four independent real functions $b_m^a+ie_m^a=c_m^a$ become
\bea
\left(\frac{1}{\rho}\pd_\rho\rho\pd_\rho-\frac{m^2}{\rho^2}+\omega^2(1-s^2)\right)b_m^1-\frac{m\mu}{\rho^2}e_m^2&=&0\label{c1}\\
\left(\frac{1}{\rho}\pd_\rho\rho\pd_\rho-\frac{m^2}{\rho^2}+\omega^2(1-s^2)\right)e_m^1+\frac{m\mu}{\rho^2}b_m^2&=&0\\
\left(\frac{1}{\rho}\pd_\rho\rho\pd_\rho-\frac{m^2}{\rho^2}+\omega^2(1-s^2)\right)b_m^2+\frac{m\mu}{\rho^2}e_m^1&=&0\\
\left(\frac{1}{\rho}\pd_\rho\rho\pd_\rho-\frac{m^2}{\rho^2}+\omega^2(1-s^2)\right)e_m^2-\frac{m\mu}{\rho^2}b_m^1&=&0\label{c4}.
\eea
\no Highly advantageously, the equations decouple into two sets of equations, which only differ by the sign of $m$. Therefore $b_m^1\sim b_{-m}^2$ and $e_m^1\sim e_{-m}^2$, and it suffices to solve the first and fourth equation of \prefr{c1}{c4}. As now the last term has the same sign, it is again possible to decouple the equations by looking for solutions of type $b_m^+=b_m^1+e_m^2$ and $b_m^-=b_m^1-e_m^2$. The original functions can be recovered by
\bea
b_m^1&=&\frac{1}{2}(b_m^++b_m^-)\nn\\
e_m^2&=&\frac{1}{2}(b_m^+-b_m^-)\nn.
\eea
\no Note that the scale constant for the two solutions $b_m^\pm$ can be selected independently: It is possible to have $b_m^1=\pm e_m^2$. The decoupled equations are then
\bea
\left(\frac{1}{\rho}\pd_\rho\rho\pd_\rho-\frac{m^2}{\rho^2}+\omega^2(1-s^2)\right)b_m^+-\frac{m\mu}{\rho^2}b_m^+&=&0\nn\\
\left(\frac{1}{\rho}\pd_\rho\rho\pd_\rho-\frac{m^2}{\rho^2}+\omega^2(1-s^2)\right)b_m^-+\frac{m\mu}{\rho^2}b_m^-&=&0.\nn
\eea
\no These equations are the same up to the sign of $m$, yielding $b_m^+=b_{-m}^-$. Note especially that for $b_m^1$ and $e_m^2$ to be convergent simultaneously, at least $b_m^+$ or $b_m^-$ has to be convergent. A cancellation of divergencies if both are divergent is not possible. Hence it finally suffices to solve the single equation
\be
\left(\frac{1}{\rho}\pd_\rho\rho\pd_\rho-\frac{m^2}{\rho^2}+\omega^2(1-s^2)\right)b_m^+-\frac{m\mu}{\rho^2}b_m^+=0.\label{bmpeq}
\ee
\no The small-$\rho$ solution for $b_m^+$ is the same as before, \pref{irsol}. At large $\rho$ the same applies as previously, but it is now more interesting to treat the zero-modes. In this case the equation at large $\rho$ is given by ($\mu(\infty)=2n+1$ for a vortex of flux $2n+1$)
\be
\left(\frac{1}{\rho}\pd_\rho\rho\pd_\rho-\frac{m^2+m(2n+1)}{\rho^2}\right)b_m^+=0\nn.
\ee
\no Since $m$ is integer, there are three possibilities.

First, if $m^2+(2n+1)m>0$, which is the case for $m>0$ or $m<-(2n+1)$, the solution is given by $\rho^{\pm\sqrt{m^2+m(2n+1)}}$. It then remains to determine whether the small-$\rho$ convergent solution connects to the large-$\rho$ convergent solution or not. It will turn out that the convergent small-$\rho$-solution connects to the divergent large-$\rho$ solution and vice versa. Thus this type of solution cannot lead to an admissible zero-mode.

Second, if $(2n+1)m=m^2$, which is possible for negative, odd $m$, the second term vanishes, and the solution is of type $c+\ln(r)$. It is again important to determine whether the small-$\rho$ convergent solution connects to the constant or divergent solution. Again it will turn out that a small-$\rho$ convergent solution connects to a large-$\rho$ divergent one.

The last possibility is $m^2+(2n+1)m<0$, which is the case for $m$ negative and $|m|<(2n+1)$. This is only possible for $n>0$, and only flux three and higher vortices can produce such solutions. This yields a large-$\rho$ behavior of
\bea
b_m^+&=_{\rho\to\infty}&C_1\cos\left(\sqrt{|m^2+m(2n+1)|}\ln(\rho)\right)\nn\\
&&+C_2\sin\left(\sqrt{|m^2+m(2n+1)|}\ln(\rho)\right).\nn
\eea
\no Although this is not vanishing, it is finite. In these cases there exist finite zero-modes, as will be confirmed below. So, $2n$ permitted absolute values of $m$ exist for each $n$. Due to the different angular structure, positive and negative $m$-solutions are independent. Hence, all in all $4n$ zero-modes for a vortex of flux $2n+1$ exist. E.\ g.\ for $n=1$ there are for the $+$-function one for $-1$ and one for $-2$. For $m=-3$, $m^2+(2n+1)m=0$. In addition, there are two for the $-$-function for $m=1$ and $m=2$. Note that for $n=0$ the vortex has a non-vanishing flux of one but does not support any zero-modes. The angular quantum number of the $r-\eta$ part of the solution has to be zero and thus does not increase the multiplicity.

Furthermore, neither the small nor the long distance behavior depends on the the specific form of $\mu(\rho)$: The existence and number of zero-modes is independent of the vortex profile, only the total flux is relevant.

It is finally necessary to obtain the intermediate range. To perform this task, equation \pref{bmpeq} has to be solved. As in the case of the instantons, this is possible using a series expansion, provided a series expansion of $\mu=\sum_i\mu_i\rho^i$ exists (which will necessarily have $\mu_0=0$). This condition is very reasonable for a smooth function going form 0 to $2n+1$ without any singularities. The solution of equation \pref{bmpeq} is then given by
\bea
b_m^+&=&\rho^{|m|}\sum_{n=0}^{\infty}b_{mn}^+\rho^n\label{bmp}\\
b_{m-1}^+&=&0\nn\\
b_{m0}^+&=&D\nn\\
b_{mn}^+&=&\frac{-\omega^2(1-s^2)b_{mn-2}^++m\sum_{i=0}^{n-1}b_{mn-1-i}^+\mu_{1+i}}{n^2+2n|m|}\nn.
\eea
\no Unfortunately, as in the instanton case, this is again a series of hyper-geometric type, and it is not practical to use it for explicit calculations for large $\rho$. This will again be remedied by use of a numerical treatment.

\subsection{Results}\label{rvor}

\begin{figure}
\epsfig{file=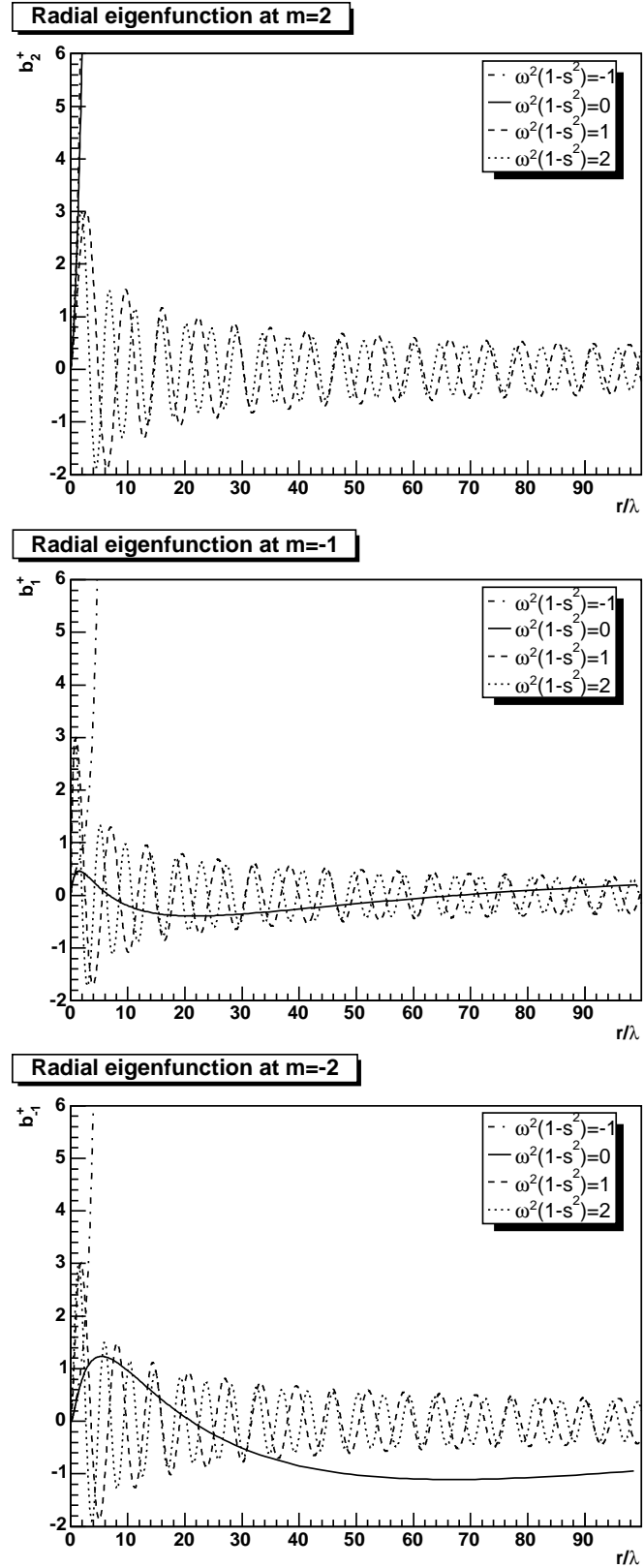,width=\linewidth}
\caption{The radial eigenfunctions $b_m^+$ for different $m$ and $\omega^2(1-s^2)$-values in a flux 3 vortex. For better visualization, positive $\omega^2$-solutions have been normalized so that their maximum is $3$, while modes with $\omega^2\le 0$ have been normalized so that $b_m^+/\rho^{|m|}|_{\rho=0}=1$.}\label{figf3}
\end{figure}

\begin{figure}
\epsfig{file=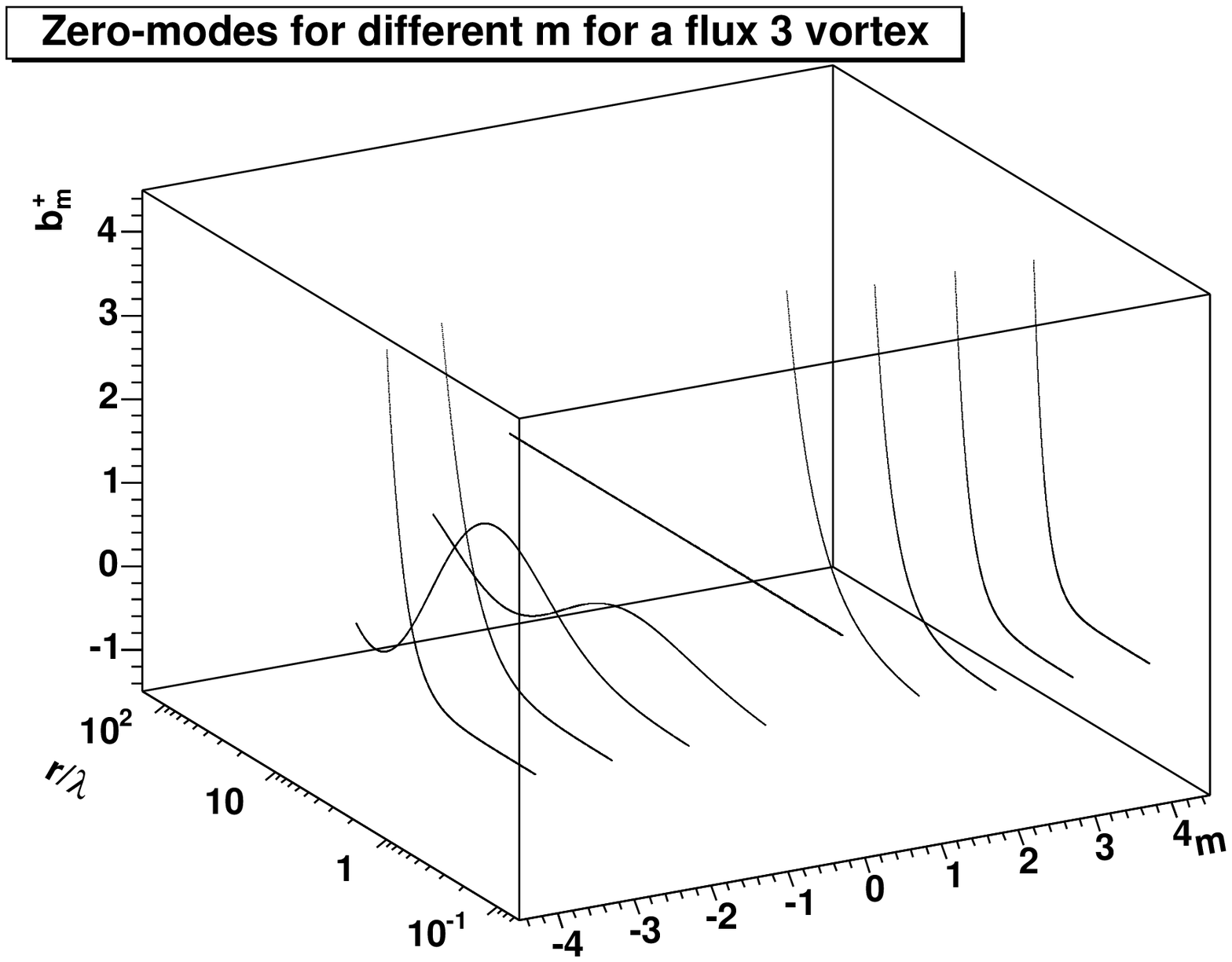,width=\linewidth}
\epsfig{file=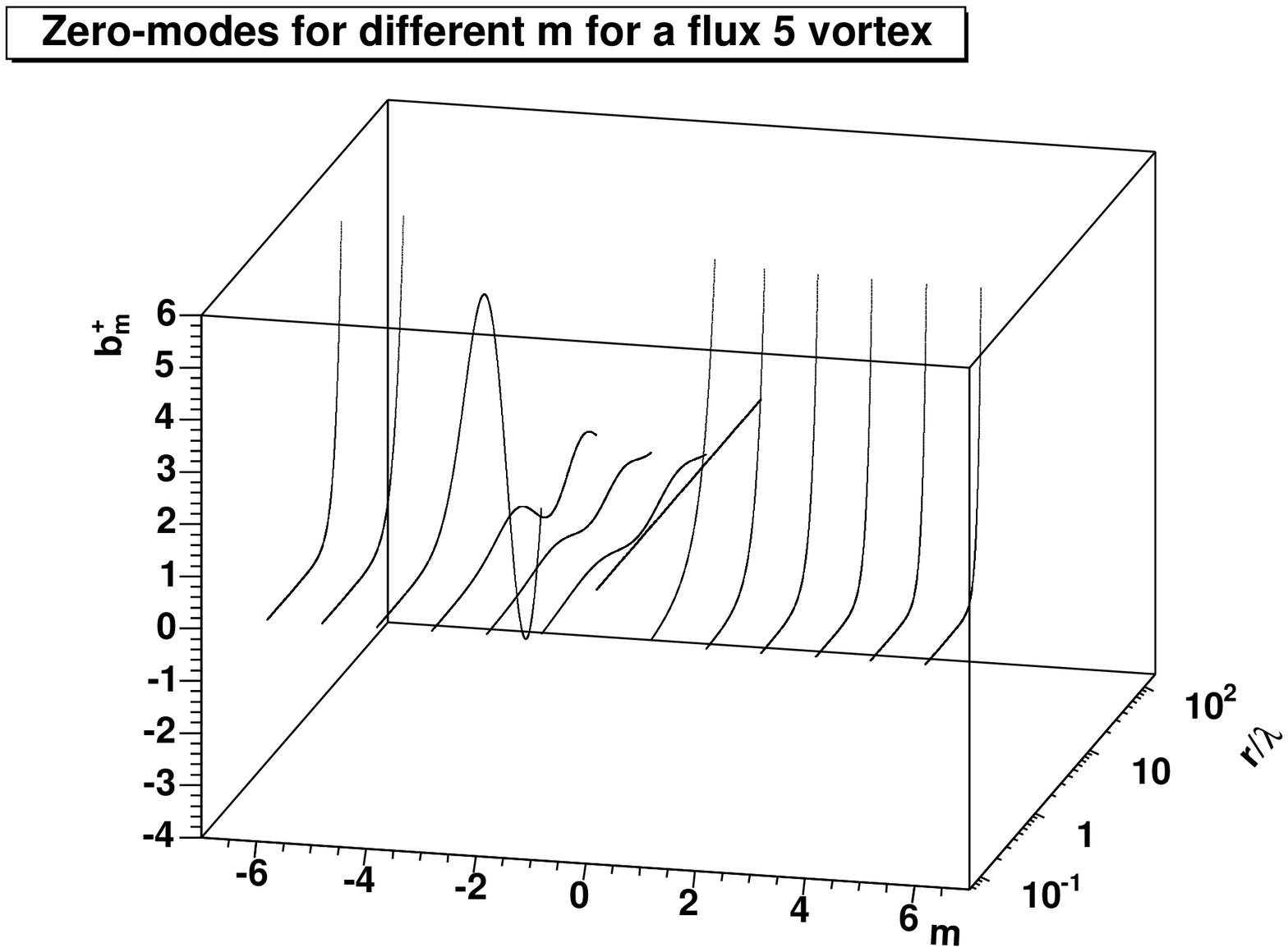,width=\linewidth}
\caption{The radial zero-modes $b_m^+$ for different $m$ in a flux 3 vortex in the top panel and in a flux 5 vortex in the bottom panel. The solutions have been normalized so that $b_m^+/\rho^{|m|}|_{\rho=0}=1$.}\label{figm1}
\end{figure}

The calculations above are confirmed by a numerical calculation, using the same techniques as in the instanton case. A flux zero vortex of profile $\mu=\rho\lambda/(\lambda^2+\rho^2)$ and a flux one vortex of profile $\mu=\rho/(\lambda+\rho)$ do not sustain zero-modes. The results for a flux three vortex and a representative selection of $m$- and $\omega^2(1-s^2)$-values are shown in figure \ref{figf3}. It is clearly visible how at negative $m$ the different convergent solutions appear, confirming the analytical calculations. This is even more evident in figure \ref{figm1}, where the zero-modes for flux three and flux five vortices with the profile $(2n+1)\rho/(\lambda+\rho)$ are shown as a function of $r$ and $m$.

Further calculations have been performed with other vortex profiles\footnote{Some of these profiles do not have a simple series expansion, so \pref{bmp} is only of a limited value. Nonetheless, these are amendable to numerical calculations, and the results do not differ qualitatively from profiles where a simple expansion exists.}, taken from \cite{Diakonov:1999gg}, e.g. $\mu=\exp(-\rho/\lambda-\lambda/\rho)$, $\mu=(2n+1)$ $\exp(-\lambda^3/\rho^3)$ and $\mu=\rho^6/(\lambda^6+\rho^6)$. These calculations do not yield qualitatively different results than the above findings.

\section{A confinement criterion}\label{ffunc}

In \cite{Greensite:2004ur} it has been argued that the criterion
\be
\lim_{\omega^2\to 0}\frac{\gamma(\omega^2)F(\omega^2)}{\omega^2}>0\label{cond}
\ee
\no is in Coulomb gauge a necessary criterion for confinement, although not sufficient. A fulfillment leads to a divergent energy of an unscreened color charge. This is of course not the ground state of QCD, but this must occur in a confining theory. The quantity $\gamma$ is the density of eigenmodes at eigenvalue $\omega^2$, and the function $F$ is the expectation value of the negative Laplace operator in the eigenstates of the Faddeev-Popov operator to eigenvalue $\omega^2$. The derivation of this condition only applies in Coulomb gauge. Thus, this criterion cannot be applied to the instanton configuration, but only to the center-vortex configuration, when the vortex is taken to be not aligned in the time direction.

In the following the criterion will be studied for the vacuum, the instanton (which will be included for completeness), and the vortex.

\subsection{Vacuum}

The eigenfunctions in the vacuum are plane waves. There are no non-trivial zero-modes, $\gamma(0)=0$ (in fact, it vanishes as $\sim\omega$ \cite{Greensite:2004ur}). Hence, it remains to determine the function $F$. In \cite{Greensite:2004ur} it is given by
\be
F(\omega^2)=-\sum_n\int d^4x\phi_{\omega^2\;n}^{a*}(x)\pd^2\phi_{\omega^2\;n}^a(x),\nn
\ee
\no where $\phi^a_{\omega^2\; n}$ is the $n$-th eigenstate to the eigenvalue $\omega^2$. In an infinite volume, this has to be normalized differently as
\be
F(\omega^2)=-\frac{\sum_n\int d^4x\phi_{\omega^2\;n}^{a*}(x)\pd^2\phi_{\omega^2\;n}^a(x)}{\sum_n\int d^4x\phi_{\omega^2\;n}^{a*}(x)\phi_{\omega^2\;n}^a(x)}.\nn
\ee
\no The corresponding calculations for the eigenfunctions\\ $C^a\exp(ik^a_\mu x_\mu)$, with $k^2=\omega^2$, can then be done explicitly to yield
\be
F(\omega^2)=\frac{k^2\sum_a|c_a|^2\int d^4x\int d^3\Omega}{\sum_a|c_a|^2\int d^4x\int d^3\Omega}=k^2\equiv\omega^2.
\ee
\no Thus, $F(\omega^2)=\omega^2$, as already stated in \cite{Greensite:2004ur}. Consequently, the condition \pref{cond} is not fulfilled, as $\gamma(\omega^2)\cdot\omega^2/\omega^2=0$ as $\omega^2\to 0$, since $\gamma(0)=0$.

\subsection{Instanton}

In cases with a non-trivial gauge field, the situation is more complicated. In both cases treated here the eigenfunctions are eigenstates of the angular momentum operator, and it would be possible (and in fact has been done) to calculate the condition explicitly, at least for $\omega^2=0$. This is not very enlightening, and a different way is pursued here. Note first that $\gamma(0)\neq 0$, as there exist for the instanton as well as for the vortex non-trivial zero-modes. Furthermore, as the states are eigenstates of the Faddeev-Popov operator, it follows that (for transverse configurations)
\be
-\phi^{a*}_{\omega^2\;n}\pd^2\phi^a_{\omega^2\;n}=\omega^2\phi^{a*}_{\omega^2\;n}\phi^a_{\omega^2\;n}+f^{abc}A_\mu^b\phi^{a*}_{\omega^2\;n}\pd_\mu\phi^c_{\omega^2\;n}.\nn
\ee
\no $F$ is then given by
\be
F(\omega^2)=\omega^2+\frac{\sum_n\int d^4x f^{abc}A_\mu^b\phi^{a*}_{\omega^2\;n}\pd_\mu\phi^c_{\omega^2\;n}}{\sum_n\int d^4x\phi_{\omega^2\;n}^{a*}\phi_{\omega^2\;n}^a},\nn
\ee
\no and the vacuum result follows directly without any calculation. It is furthermore clear that if the gauge field and thus the eigenmodes are regular everywhere, the second term has to be behave as $-\omega^2$ for $\omega^2\to 0$ in order to prevent the fulfillment of \pref{cond} in case of a non-zero level-density at $\omega^2=0$.

In the instanton case, the eigenmodes are eigenfunctions of $f^{abc}(x^2+\lambda^2)A_\mu^b\pd_\mu$. Using then the orthogonality of the angular part of the eigenfunctions to perform the angular integrations yields
\be
F(\omega^2)=\omega^2+\frac{\sum_n\int x^3dx f^{abc}\frac{4c_n}{x^2+\omega^4}\phi^a_{\omega^2\;n}\phi^a_{\omega^2\;n}}{\sum_n\int x^3dx\phi_{\omega^2\;n}^a\phi_{\omega^2\;n}^a}\nn
\ee
\no With the known exact formulas for the radial part, the integrations can be performed exactly, but yield a lengthy expression involving several poly-logarithms. This expression, however, vanishes in the infinite volume limit, in accordance with a numerical integration. In this case $F(\omega^2)=\omega^2$, as in the vacuum. Although now \pref{cond} is fulfilled, nothing is implied by this, as the instanton is not a Coulomb gauge configuration. Still, this result is not quite the one which would be naively expected for an instanton, but it is not so suprising when considering the presence of zero-modes.

\subsection{Vortices}

Different Fourier modes are not coupled by the gauge field. Therefore, it follows that
\bea
&&f^{abc}\frac{\delta^{b3}\mu(\rho)}{\rho^2}\phi_{\omega^2\;mn}^{a*}\pd_\eta\phi_{\omega^2\;mn}^{c}\nn\\
&=&f^{a3c}\frac{\mu(\rho)}{\rho^2}\phi_{\omega^2\;mn}^{a*}\pd_\eta\phi_{\omega^2\;mn}^{c}\nn\\
&=&\frac{im\mu{\rho}}{\rho^2}\left(\phi_{\omega^2\;mn}^{2*}\phi_{\omega^2\;mn}^{1}-\phi_{\omega^2\;nm}^{1*}\phi_{\omega^2\;mn}^{2}\right)\nn\\
&=&\frac{2m\mu(\rho)}{\rho^2}\left(b_{\omega^2\;mn}^1e_{\omega^2\;mn}^2-b_{\omega^2\;mn}^{2}e_{\omega^2\;mn}^1\right).\nn
\eea
\no In the last line the notation of the previous calculations has been used. Such contributions exist for all allowed $m$ values. As the eigenfunctions are independent, but have a common long range behavior, the fraction in $F$ only vanishes if the denominator scales with a larger power of the volume than the numerator. Otherwise any desired result could be obtained by an appropriate choice of the independent normalization constants. In the denominator this is of no importance, as the integrated functions are positive semidefinite. Thus, no singularity can be constructed. In addition, from the trivial two dimensions, each term is multiplied by a factor stemming from a free-wave solution, i.\ e.\ a term proportional to the square root of the volume, which cancels out.

The zero-modes all behave essentially similar to
\be
\frac{\rho^{|m|}}{\lambda^{|m|}+\rho^{|m|}}\cos\left(\sqrt{|m^2+m(2n+1)|}\ln\left(\frac{\rho+\lambda}{\lambda}\right)\right),\nn
\ee
\no where the scale $\lambda$ is set by the function $\mu$. Numerical integration shows (when $\mu$ is majorized by a constant) that the numerator scales like the logarithm of the volume while the denominator scales as the square root of the volume. This is as anticipated due to the explicit factor of $1/\rho^2$.

Thus, independent of the selected normalization constants, the second term in $F$ vanishes like $\ln(V)/V^{1/2}$, leaving again only $F(\omega^2)=\omega^2$. Since for vortices with fluxes higher than one the level-density does no longer vanish at zero, the condition \pref{cond} is fulfilled. As these are Coulomb gauge configurations, this necessary confinement criterion is met.

It should be noted that the result is not the same as on a finite lattice in a thin-center-vortex-only configuration \cite{Greensite:2004ur}, which seems to indicate a finite value of $F(0)$. However, this depends significantly on the volume, and from the available ones, it cannot be excluded that $F(0)$ does not vanish in the infinite-volume limit also in lattice calculations.

\section{Discussion and conclusions}\label{dc}

Summarizing, the spectrum of the Faddeev-Popov operator in a one-instanton background field has the following features: It is positive semi-definite. It has a continuous spectrum of positive eigenvalues, just as in the vacuum, but with a different angular structure. The (infinite) degeneracy of non-zero eigenvalues does not depend on the eigenvalue, and there is no enhancement of eigenmodes near zero, but only at zero. Three new zero-modes exist at ``angular momentum'' $l=1/2$ and $l=1$. This number is independent of the instanton size $\lambda$, which is the only characteristic quantity of an instanton. Thus the instanton belongs to the first Gribov horizon, the boundary of the first Gribov region.

Furthermore, the configuration is due to \pref{minfunc} also part of the fundamental modular region, and lies on the common boundary of the first Gribov horizon and the fundamental modular region. Hence, it is contained within the region of field configuration space which is responsible for gluon confinement according to the Gribov-Zwanziger scenario. Still, the one instanton field does not lead to a significant enhancement of the number of zero-modes, as required by the Gribov-Zwanziger scenario. It is therefore likely not alone, if at all, responsible for or even involved in gluon confinement.

Still, it is tempting to speculate whether an $n$-instanton configuration, consisting of far separated instantons, could be able to support $n$ independent zero-modes. The radial structure supports this, as at sufficiently far distances radial eigenmodes behave like the vacuum solution or vanish. But the intricate angular variations, which still exist at infinity, makes it far from obvious that a multi-instanton configuration does sustain more zero-modes. Only the $l=1/2$ solution shows any sign of localization, and may thus be a candidate for a relevant multi-instanton solution.

Hence, it cannot yet be completely concluded that instantons are irrelevant to gluon confinement. Therefore, it would be interesting to study this connection also in lattice gauge theory. This is, however, made complicated due to the appearance of sub-structure of topological defects on toroidal manifolds, which has been seen e.\ g.\ in connection with Kraan-van Baal calorons \cite{Kraan:1998pm} in lattice calculations \cite{Kraan:1998sn}. Digressing, this rises the very interesting question, if in these cases any influence on the eigenspectrum of the Faddeev-Popov operator will be due to the constituents or will arise as a feature of the bound states.

On the other hand, a vortex of sufficiently high flux can sustain an arbitrary number of zero-modes. However, a vortex of flux one is not sufficient for this; a larger flux is necessary. Only flux three or higher vortices belong to the Gribov horizon, while vortices with lower flux are only part of the interior of the first Gribov region. Thus, vortex configurations, if they belong to the fundamental modular region\footnote{As vortex configurations appear as physical excitations in lattice calculations \cite{Bertle:2000py}, this seems likely.}, could contribute to the enhancement of the spectrum of the Faddeev-Popov operator, and even be its main source. This is again only an enhancement at eigenvalue zero, as the (infinite) degeneracy of non-zero modes is again not dependent on the eigenvalue.

 The fact that a single vortex can sustain a large number of zero-modes is different from the instanton case, where a single instanton was not able to support more than three zero-modes, irrespective of its size. This is to some extent surprising, as it has been shown that a vortex does not carry topological charge when it is oriented \cite{Reinhardt:2001kf}, as it is the case in the present calculation. The only relevant property seems to be a sufficiently large flux.

It is noteworthy that the zero-modes, with the exception of the $l=1/2$ one in the instanton case, are not localized, i.\ e.\ do not vanish at spatial infinity. Especially the vortex zero-modes do still vary appreciably even in radial direction at spatial infinity. This indicates that eventually non-localized modes may be important in the confinement problem. This is supported by similar observations in recent lattice investigations of the covariant Laplacian \cite{Greensite:2005yu} and also of the Faddeev-Popov operator \cite{Sternbeck:2005vs}. Also in the context of the Gribov-Zwanziger scenario, the non-localization of such modes play a role \cite{Zwanziger:2003cf}. Still this issue is not finally settled, and it is not yet really clear, what the significance of localization in this context is.

In addition, the vortex has been found to satisfy one necessary criterion for quark confinement in Coulomb gauge. Thus, if all of these indications are correct, and such vortices play a significant role in the dynamics of Yang-Mills theory and eventually of QCD, they are perhaps the link searched for in this work. Nonetheless, it is likely that vortices are still only a symptom rather than the origin of confinement, as confinement exists also in Yang-Mills theories with centerless gauge groups like G$_\mathrm{2}$ \cite{Holland:2003jy}. In these cases, possibly different types of topological excitations play the same role as vortices in SU(2).

Concluding, the following results have been found: Instantons belong to the common boundary of the first Gribov region and the fundamental modular region, but only support a small number of zero-modes, independent of the size of the instanton. This shows that not all configurations on the common boundary provide a significant enhancement of the spectrum of the Faddeev-Popov operator at eigenvalue zero. Instantons seem again not directly relevant to confinement, but it is not clear whether an ensemble of instantons can sustain more zero-modes.

Vortices of flux zero and also of flux one do not belong to the common boundary nor even to the Gribov horizon and therefore do not play any role in the Gribov-Zwanziger scenario. This implies that not all topological field configurations lie on the boundary nor provide zero-modes in the spectrum of the Faddeev-Popov operator. Vortices of flux $2n+1\ge 3$ belong to the Gribov horizon, and lie always in the first Gribov region, although pointwise $|A_\mu|$ is not bound for large $n$. (Oriented) vortices can support $4n$ zero-modes, and thus provide a significant enhancement of the spectrum at zero of the Faddeev-Popov operator. This corresponds to results in lattice calculation that the enhancement of the spectrum of the Faddeev-Popov operator vanishes when the vortex content is removed \cite{Greensite:2004ur}. Hints for a related modification of the propagator have been found on the lattice \cite{Langfeld:2002dd}. This indicates that perhaps the ghosts are the place to look for traces of topological excitations in functional methods.

Finally, these results demonstrate that certain types of topological configuration influence the spectrum of the Faddeev-Popov operator, and can give rise to the appearance of zero-modes. Furthermore, it has been shown that this strongly depends on the type of configuration, and that even some configurations do not provide an enhancement at eigenvalue zero. These results provide a first step towards an analytical understanding of the intimate link apparently existing between a confinement dominated by topological aspects and the Gribov-Zwanziger scenario, which are both by now well supported by various calculations. In addition, the results presented here also support similar findings in numerical lattice calculations. 

\acknowledgement

The author is grateful to Reinhard Alkofer and Jan M.\ Pawlow\-ski for valuable and inspiring discussions, and to Jan M.\ Paw\-low\-ski also for a critical reading of and helpful comments on this manuscript. This work was in part supported by the DFG under grant number MA 3935/1-1.

\end{document}